\documentclass[preprint,showpacs,aps,prd]{revtex4}
\usepackage{amsmath}
\usepackage{amssymb}
\usepackage{graphicx}
\usepackage{graphics}

\input{tcilatex}

\begin{document}

\title{Estimate of the energy of vacuum fluctuations of non-Abelian gauge fields from the
uncertainty relations}
\author{S. Stepanow}
\affiliation{Martin-Luther-Universit\"{a}t Halle-Wittenberg, Institut f\"{u}r Physik,
D-06099 Halle, Germany}
\date{\today}

\begin{abstract}
We derive the commutation relations for field strengthes of the non-Abelian gauge fields by requiring that the quantum mechanical equations of motion coincide with the classical field equations. The equations of motion with respect to time derivative coincide with the corresponding field equations, while those with respect to space derivatives agree with classical field equations if constraint equations are fulfilled.  Using the uncertainty relations for field strengthes at the same times, which follows from the commutation
relations, we estimate the energy gap of vacuum fluctuations in the long wave limit
as $\mathcal{E}_{0,k} \sim  c\hbar\Lambda^{-2} V^{-1}
   \left(g^2/c\hbar\right)^{1/3}$ and the total energy $\mathcal{E}_0\sim \Lambda c\hbar\left( g^{2}/\hbar c\right) ^{1/3}$
with $\Lambda $ being a cutoff separating the weak and strong coupling regimes.
%small $k$ from the large ones.
\end{abstract}

\pacs{03.70.+k,11.15.-q,12.38.Aw}
\maketitle

%\email[Corresponding author: email\ ]{stepanow@physik.uni-halle.de}

%03.70.+k 	Theory of quantized fields (see also 11.10.-z Field theory)
%11.15.-q 	Gauge field theories
%11.15.Tk 	Other nonperturbative techniques
%12.38.Aw 	General properties of QCD (dynamics, confinement, etc.)
%12.38.Lg 	Other nonperturbative calculations

The non-Abelian gauge fields (Yang-Mills fields) \cite{yang-mills54}, which  are introduced by
generalization of the local gauge symmetry of the Maxwell electrodynamics from $%
U(1)$ to $SU(n)$, are the basis of the unified theory of electro-weak
interactions ($U(1)\times SU(2)$), and strong interactions ($SU(3)$). The
quantization of these theories, which is available by now in the literature only in the perturbational limit, involves peculiarities and demands an introduction of ghost fields to fulfill the unitarity condition \cite{feynman},\cite{faddeev-popov67},\cite{dewitt}.
%The Yang-Mills fields are non linear, and the analytical predictions are scarce.

In the present Letter we will perform the quantization of the Yang-Mills fields by requiring that the Heisenberg's (quantum mechanical) equations of motion coincide with the classical field equations. This approach is closely related to Dirac's method of quantization of a classical system, which consists in appropriate replacement of the Poisson brackets by the commutation relations. The Heisenberg's equations of motion and their generalization to the space derivatives (see Eq.~(\ref{hbg-eb-r}) for
electromagnetic field) can be regarded as the differential (local)
formulation of the unitarity condition similar to the quantum mechanical systems
with a few degrees of freedom. Using the uncertainty relations for electric
and magnetic field strengthes, which follows from the corresponding
commutation relations, we estimate the ground state energy of the Yang-Mills
fields in the long wave limit as $\mathcal{E}_{0,k} \sim  c\hbar\Lambda^{-2} V^{-1}
   \left(g^2/c\hbar\right)^{1/3}$, which depends non
analytically on the fine structure constant, and on a microscopic cutoff $\Lambda $, which separates the strong and weak coupling limits. In the method used in this Letter the nonlinear quantum dynamics of Yang-Mills fields is expressed by the simultaneous commutation relations, which allow one to derive using the uncertainty relations the expectation values for vacuum fluctuations of field strengthes by fully taking into account the non linearities.
Our estimate using the uncertainty relations can be considered as a simplest version of the Rayleigh-Ritz method of quantum mechanics, and is similar from the technical side to the estimate of the ground state energy of interacting oscillators. We expect that despite the rather crude derivation the above result is robust and represent a lower bound for the vacuum energy of the Yang-Mills fields.

To elucidate the procedure we will first consider the quantization of the
free electromagnetic field via Heisenberg's equations of motion \cite%
{dirac-lectures}, and the estimate of the vacuum energy using the
uncertainty relations. The requirement that the quantum mechanical equations
of motion for field strengths
\begin{equation}
\frac{d\mathbf{E}}{dt}=\frac{i}{\hbar }\left[ \mathcal{H},\mathbf{E}\right]
,\ \ \ \frac{d\mathbf{B}}{dt}=\frac{i}{\hbar }\left[ \mathcal{H},\mathbf{B}%
\right] ,  \label{hbg-eb}
\end{equation}%
where
\begin{equation}
\mathcal{H}=\frac{1}{2}\int d^{3}r\left( \mathbf{E}^{2}(r)+\mathbf{B}%
^{2}(r)\right)   \label{H-EB}
\end{equation}%
is the Hamilton operator of the free electromagnetic field in
Lorentz-Heaviside units, coincide with the Maxwell equations
\begin{equation}
\frac{\partial \mathbf{E}}{\partial t}=\mathrm{rot}\mathbf{B},\ \ \ \frac{%
\partial \mathbf{B}}{\partial t}=-\mathrm{rot}\mathbf{E},  \label{rot-EB}
\end{equation}%
requires the fulfilment of the following commutation relations for field
strengths
\begin{eqnarray}
\left[ B^{l}(\mathbf{r}^{\prime },t),E^{n}(\mathbf{r},t)\right]  &=&i\hbar
\varepsilon ^{lnm}\frac{\partial }{\partial x^{\prime m}}\delta (\mathbf{r}%
^{\prime }-\mathbf{r}),  \notag \\
\left[ E^{l}(\mathbf{r}^{\prime },t),E^{n}(\mathbf{r},t)\right]  &=&0,
\notag \\
\left[ B^{l}(\mathbf{r}^{\prime },t),B^{n}(\mathbf{r},t)\right]  &=&0.
\label{vr_EB}
\end{eqnarray}%
The derivatives $d/dt$ in Heisenberg's equations of motion have been
replaced by partial derivatives $\partial /\partial t$ in Maxwell equations.
The Heisenberg's equations of motion (\ref{hbg-eb})
%, which are differential equations with respect to time,
can be considered as the differential form of the unitarity condition ensuring the conservation of probability during the time evolution of the system.

The equations of motion with respect to space derivatives are given by
\begin{equation}
\frac{\partial E^{m}}{\partial x^{i}}=\frac{-i}{\hbar }\left[ \mathcal{P}%
^{i},E^{m}\right] ,\ \frac{\partial B^{m}}{\partial x^{i}}=\frac{-i}{\hbar }%
\left[ \mathcal{P}^{i},B^{m}\right] ,  \label{hbg-eb-r}
\end{equation}%
where
\begin{equation*}
\boldsymbol{\cal P}=\int d^{3}r\mathbf{E}(\mathbf{r},t)\times \mathbf{B}(%
\mathbf{r},t)
\end{equation*}%
is the momentum of the electromagnetic field. The explicit calculation of
the commutators in Eqs.~(\ref{hbg-eb-r}) using the commutation relations (%
\ref{vr_EB}) shows that the quantum mechanical equation of motion (\ref%
{hbg-eb-r}) are equivalent to the Maxwell equations
%\cite{stepanow}
\begin{equation}
\mathrm{div}\mathbf{E}=0,\ \ \mathrm{div}\mathbf{B}=0.  \label{div-EB}
\end{equation}
The equations of motion (\ref{hbg-eb-r}) can be considered similar to (\ref{hbg-eb}) as the differential form of the unitarity condition ensuring the conservation of the probability by space propagation of the electromagnetic field.

To estimate the vacuum fluctuations of the quantized electromagnetic field
using the commutation relations (\ref{vr_EB}) we follow the consideration
for a harmonic oscillator and replace $p$ in the energy $E=p^{2}/2+\omega ^{2}q^{2}/2$
according to $p\simeq \hbar /2q$, which follows
from the commutator $[q,p]=i\hbar $, and results after minimization in the exact expression for
the ground state energy $E_{0}=\hbar \omega /2$. For the electromagnetic
field ones uses the uncertainty relations between the strengthes of the
electric and magnetic field for a mode propagating in $z$-direction in the
form $\overline{(\Delta B_{k}^{x})^{2}}\times \overline{(\Delta
E_{-k}^{y})^{2}}$ $\simeq (\hbar k^{z}/2V)^{2}$. The elimination of the strengthes of the magnetic field from the field energy (\ref{H-EB}) and
minimization with respect to $\overline{E_{k}E_{-k}}$ one obtains $\overline{%
E_{k}E_{-k}}$ $=\hbar k/2V$, and finally $2\times \hbar \omega _{k}/2$ for
the vacuum energy, where the factor $2$ takes into account the number of
polarizations.

The Yang-Mills fields are introduced by requiring the invariance of the Lagrangian with respect to the local
gauge transformations of the matter field, which in the case of $SU(2)$
consists of two components, and transforms according to $\psi =\exp (-ig\theta
^{a}T^{a})\psi ^{\prime }$ (in relativistic units $\hbar =c=1$) with $%
T^{a}=\sigma ^{a}/2$, and $\sigma ^{a}$ being the Pauli matrices. The local gauge invariance demands a
coupling to an external Yang-Mills field, which is introduced by replacing $%
\partial _{\mu }$ with the covariant derivative (see for example \cite%
{ytzykson-zuber}) according to
\begin{equation*}
\partial _{\mu }\rightarrow D_{\mu }=\partial _{\mu }+igW_{\mu },
\end{equation*}%
where the shortcut $W_{\mu }=W_{\mu }^{a}T^{a}$ is used. The sum convention
over the isotopic index $a$ is implied above and in the following. The gauge
invariant Lagrangian of the Yang-Mills field is given by \cite%
{ytzykson-zuber}
\begin{equation*}
\mathcal{L}=-\frac{1}{4}G_{\mu \nu }^{a}G_{a}^{\mu \nu },
\end{equation*}%
where the components of the tensor of field strengthes are given by
\begin{equation}
G_{\mu \nu }^{a}=\partial _{\mu }W_{\nu }^{a}-\partial _{\nu }W_{\mu
}^{a}-g\varepsilon ^{abc}W_{\mu }^{b}W_{\nu }^{c}.  \label{fstens}
\end{equation}%
The field strengths are expressed through $G_{\mu \nu }^{a}$ in the same way
as in electrodynamics
\begin{equation*}
E_{a}^{k}=G_{a}^{k0},\ B_{a}^{k}\ =-\frac{1}{2}\varepsilon ^{klm}G_{a}^{lm}, G^{lm}_a=-\varepsilon^{lmk}B^k_a.
\end{equation*}%
The second pair of the classical field equations can be obtained from the
Euler-Lagrange equations
\begin{equation*}
\frac{\partial \mathcal{L}}{\partial W_{\sigma }^{a}}-\frac{\partial }{%
\partial x^{\gamma }}\frac{\partial \mathcal{L}}{\partial W_{\sigma ,\gamma
}^{a}}=0
\end{equation*}%
with $W_{\sigma ,\gamma }^{a}=\partial W_{\sigma }^{a}/\partial x^{\gamma }$
as $D^{\nu }G_{\mu \nu }=0$, and can be written for components $G_{\mu \nu }^{a}$ as
\begin{equation}
\partial _{\nu }G_{a}^{\mu \nu }+g\varepsilon _{abc}G_{b}^{\mu \gamma
}W_{\gamma }^{c}=0.  \label{1pair}
\end{equation}%
The isotopic indices can be written in equivalent way as subscripts or superscripts. The
first pair of the field equations (Bianchi identity) follows from the
relation between the field strengthes and the vector potentials given by Eq.~(\ref{fstens}), and reads
\begin{equation*}
D^{\sigma }G^{\mu \nu }+D^{\mu }G^{\nu \sigma }+D^{\nu }G^{\sigma \mu }=0.
\end{equation*}%
The latter result in the following explicit equations for $G_{a}^{\mu \nu }$
\begin{eqnarray}
&&\partial ^{\sigma }G_{a}^{\mu \nu }+\partial ^{\mu }G_{a}^{\nu \sigma
}+\partial ^{\nu }G_{a}^{\sigma \mu }+g\varepsilon _{abc}G_{b}^{\mu \nu
}W_{c}^{\sigma }  \notag \\
&&+g\varepsilon _{abc}G_{b}^{\nu \sigma }W_{c}^{\mu }+g\varepsilon
_{abc}G_{b}^{\sigma \mu }W_{c}^{\nu }=0.  \label{2pair}
\end{eqnarray}%
The energy $T^{00}$ and the momentum $T^{0k}$ of the Yang-Mills field are given by \cite{ytzykson-zuber}
\begin{equation}
T^{00}=\frac{1}{2}\int d^{3}r\left(
E_{a}^{k}E_{a}^{k}+B_{a}^{k}B_{a}^{k}\right) ,  \label{T00}
\end{equation}%
\begin{equation}
T^{0k}=\int d^{3}r\varepsilon ^{klm}E_{a}^{l}B_{a}^{m}=\int
d^{3}rG_{a}^{i0}G_{a}^{ik}.  \label{T0k}
\end{equation}

We now will in quantizing the Yang-Mills fields require that the Heisenberg's equations of motion coincide with the classical field equations. This procedure is equivalent to the Dirac's method \cite{dirac-qm} consisting in replacing the Poisson brackets by commutation relations according to $\{A,B\}\rightarrow -i/\hbar \lbrack \hat{A},\hat{B}]$. The field strengths $G_{a}^{\mu\nu }$ and vector potentials $W_{a}^{\mu }$ are operators, which generally
do not commutate with each others, and are supposed to obey the equations (%
\ref{1pair}-\ref{2pair}). In that case the products $G_{a}^{\mu \nu
}W_{b}^{\mu }$ should be replaced by $(1/2)(G_{a}^{\mu \nu }W_{b}^{\mu
}+W_{b}^{\mu }G_{a}^{\mu \nu })$. However, we will not do this explicitly,
because our result on the estimate of the energy of vacuum fluctuations will not be affected.
In order to establish the commutation relations for the field strengthes we
compare the Heisenberg's equation of motion
\begin{eqnarray}\label{eq_Etime1}
\frac{d E_{a}^{m}}{dt} &=&\frac{i}{\hbar }\left[ T^{00},E_{a}^{m}\right]  \nonumber\\
&=&\frac{i}{\hbar }\int d^{3}r^{\prime }E_{b}^{k}(r^{\prime })\left[
E_{b}^{k}(r^{\prime }),E_{a}^{m}(r)\right]  \nonumber \\
&+&\frac{i}{\hbar }\int d^{3}r^{\prime }B_{b}^{k}(r^{\prime })\left[
B_{b}^{k}(r^{\prime }),E_{a}^{m}(r)\right]
\end{eqnarray}%
with the field equation
\begin{equation}\label{eq_Etime2}
\partial ^{0}G_{m0}^{b}+\partial ^{k}G_{mk}^{b}-g\varepsilon
_{bac}G_{m\gamma }^{a}W_{c}^{\gamma }=0.
\end{equation}%
The requirement that both equations coincide yields the commutation
relations between the different isotopic components of the strengths of the
electric field
\begin{equation}
\left[ E_{a}^{k}(r^{\prime }),E_{b}^{m}(r)\right] =-i\hbar g\varepsilon
_{abc}\delta ^{km}W_{c}^{0}\delta (r^{\prime }-r),  \label{cr_ee}
\end{equation}%
and the commutation relation between the strengthes of the electric field
and magnetic field
\begin{eqnarray}\label{cr_eb1}
    \left[B^n_a(r'),E^{l}_b(r)\right]&=&i\hbar\delta_{ab}\varepsilon^{nlm}\partial^{'m}\delta(r'-r)\nonumber \\
    &-&i\hbar g \varepsilon_{abc}\varepsilon^{nlm}W^m_c\delta(r'-r).
\end{eqnarray}
The first term on the right-hand side of (\ref{cr_eb1}) coincides with that for the
electromagnetic field (\ref{vr_EB}).

To derive the commutation relations for the strengthes of the magnetic field
we consider the equation of motion
\begin{eqnarray}
\frac{dB_{a}^{m}}{dt} &=&\frac{i}{\hbar }\left[ T^{00},B_{a}^{m}\right]  \nonumber \\
&=&\frac{i}{\hbar }\int d^{3}r^{\prime }E_{b}^{k}(r^{\prime })\left[
E_{b}^{k}(r^{\prime }),B_{a}^{m}(r)\right]  \nonumber \\
&+&\frac{i}{\hbar }\int d^{3}r^{\prime }B_{b}^{k}(r^{\prime })\left[
B_{b}^{k}(r^{\prime }),B_{a}^{m}(r)\right]
\end{eqnarray}%
and demands that the latter coincides with the field equations
(\ref{2pair}) for $\sigma =0$, $\mu =i$, $\nu =m$
\begin{eqnarray}
\frac{\partial }{\partial t}G_{a}^{im} &=&-\partial ^{i}G_{a}^{m0}-\partial
^{m}G_{a}^{0i} -g\varepsilon _{abc}G_{b}^{im}W_{c}^{0} \nonumber \\
&-&g\varepsilon
_{abc}G_{b}^{m0}W_{c}^{i}-g\varepsilon _{abc}G_{b}^{0i}W_{c}^{m}.
\end{eqnarray}%
The latter requirement will be fulfilled, if in addition to (\ref{cr_eb1}) the
commutation relation between the strengthes of the magnetic field takes the shape
\begin{equation}\label{cr_bb1}
    \left[B^k_a(r'),B^n_b(r)\right]=-i\hbar g \varepsilon_{abc}\delta^{kn}W^0_c\delta(r'-r).
\end{equation}

The evaluation of the equation of motion
\begin{equation}\label{eq-Espace1}
\partial ^{m}G_{b}^{i0}=-\frac{i}{\hbar }\left[ T^{0m},G_{b}^{i0}\right]
\end{equation}%
using the above commutation relation results in
\begin{eqnarray}\label{eq-Espace2}
    \partial ^m G^{i0}_b&=&-\partial_lG^{l0}_b\delta^{mi}+\partial^m G^{i0}_b -g\varepsilon_{abc}G^{i0}_aW^m_c \nonumber \\
    &-&g\varepsilon_{abc}\delta^{mi}G^{l0}_aW^l_c -g\varepsilon_{abc}G^{im}_aW^0_c.
\end{eqnarray}
The part of the latter which is proportional to $ \delta^{mi}$ results in the field equation
\begin{equation}\label{eq-Espace3}
    \partial_lG^{l0}_b+g\varepsilon_{bac}G^{m0}_aW^m_c=0,
\end{equation}
whereas the remainder yields the constraint equation
\begin{equation}\label{constr_E}
    \varepsilon_{abc}(G^{i0}_bW^m_c-G^{im}_bW^0_c)=0.
\end{equation}
Similarly, the equation of motion
 \begin{equation}\label{eq-Bspace1}
    \partial^mB^m_b=-\frac{i}{\hbar}\left[T^{0m},B^m_b \right],
\end{equation}
decomposes in the field equations
\begin{equation}\label{eq-Bspace2}
    \partial^mB^m_b=g\varepsilon_{abc}B^m_aW^m_c,
\end{equation}
which are associated with terms $ \sim \delta^{mi}$,
and the constraint equation
\begin{equation}\label{constr_B}
\varepsilon_{abc}(B^i_b W_c^m-\varepsilon ^{imn}E^n_bW^0_c)=0.
\end{equation}
Therefore, the equations of motion (\ref{eq-Espace1},\ref{eq-Bspace1}) coincide with the corresponding field equations only if the constraint equations (\ref{constr_E},\ref{constr_B}) are fulfilled.
The quantization via equations of motion enables one to reduce the complicated quantum dynamics of the Yang-Mills field to the commutation relations of the field strengthes (\ref{cr_ee},\ref{cr_eb1},\ref{cr_bb1}) at the same time. The constraint equations reflect the peculiarities of quantization of Non-Abelian gauge fields.

In order to check the consistency of the commutation relations (\ref{cr_ee},\ref{cr_eb1},\ref{cr_bb1}) we compare the number of commutation relations with the number of vector potentials. The total number of commutation relations (\ref{cr_ee},\ref{cr_eb1},\ref{cr_bb1}) is equal to $9+4\times 9+9$, while the number of
vector potentials is $12$.
The commutation relations between the components of the electric and magnetic field and consequently the uncertainties relations of these components are determined according to (\ref{cr_ee},\ref{cr_bb1}) by the zero component of the vector potential $W_{c}^{0}$. Using similar to the case of the electromagnetic field \cite{landau-lifschitz-bd4} the particular gauge $W_{a}^{0}=0$ has the consequence that there are no uncertainties between different isotopic components of the electric and magnetic field strengthes. This choice of gauge decreases the number of constraints. Assuming the independence of the expectation values of field strengthes in the ground state on the isotopic and the Cartesian indices one decreases further the number of constraints. As a result the over-specification of field strengthes will be lifted.

It is known that the naive canonical quantization of non-Abelian gauge fields is in conflict with the unitarity condition, which makes necessary an introduction of the  ghost fields \cite{feynman},\cite{faddeev-popov67},\cite{dewitt}.
The fixation of gauge to avoid the multiple counting of the field states differing only by the gauge occurs by the introduction of Dirac's delta functions under the path integral for the vacuum-vacuum transition amplitude, which are the source of the ghost fields. The peculiarity of quantization via equations of motion manifests itself in the constraint equations (\ref{constr_E},\ref{constr_B}). It is beyond the scope of this Letter to discuss the connection between both approaches.
The constraint equations (\ref{constr_E},\ref{constr_B}) vanish for expectation values, so that they do not affect the treatment of vacuum fluctuations in the following, which is based on the uncertainty relations.

The commutation relation (\ref{cr_eb1}) is the start point to estimate the
vacuum fluctuations of the energy of the non-Abelian gauge fields, which we will perform here in the long wave limit. It is convenient to use for this goal the commutation relations in $k$-space.
For example, the relation between the field strengthes and the vector potentials reads
\begin{eqnarray}
G_{\mu \nu }^{a}(k) &=&ik_{\mu }(W_{\nu }^{a}(k)-ik_{\nu }W_{\mu }^{a})
\notag \\
&-&g\varepsilon ^{abc}\sum_{k^{\prime }}W_{\mu }^{b}(k^{\prime })W_{\nu
}^{c}(-k-k^{\prime }).  \label{G-W}
\end{eqnarray}%
We now will derive using the commutation relations the uncertainty relations for field strengthes. In considering the limit of small wave vectors we omit the $k$-dependent terms in commutation relations. As we
have mentioned above there are no uncertainties between the different isotopic components of field strength, which is the consequence of the choice of the gauge $W_{a}^{0}=0$. The uncertainty relations between the
field strengthes of the magnetic and electric fields are obtained from the commutation relations (\ref{cr_eb1}) for small $k$ as follows
\begin{equation}
H_{a}^{3}(k^{\prime })E_{b}^{1}(k)=\frac{\hbar }{2V}g\left\vert \varepsilon
_{abc}\right\vert W_{c}^{2}(k+k^{\prime }),  \label{uncernt}
\end{equation}%
and cyclic permutations over the cartesian (upper) indices. The field
strength $H_{a}^{3}(k^{\prime })$ in (\ref{uncernt}) and in the following
formulas means $H_{a}^{3}(k^{\prime })\equiv \left( \overline{%
H_{a}^{3}(k^{\prime })^{2}}\right) ^{1/2}$, and analogous for the strengthes
of the electric field and vector potentials, where the overbar denotes the
quantum mechanical average. The elimination of the strengthes of the
magnetic field in the energy (\ref{T00}) using (\ref{uncernt}) yields after the minimization with respect to the strengthes of the electric field
\begin{equation}
\mathcal{E}_0 \simeq \frac{9}{2}\hbar g\sum_{k}W_{\mu }^{b}(k),  \label{energ}
\end{equation}%
where we assumed that the expectation values are independent of the cartesian and
isotopic indices \footnote{Note that in performing the minimization we neglect the relation between the field strengthes and the vector potential. }.
The factor $9=3\times 3$ apply for $SU(2)$. The
elimination of the expectation value of the potential in (\ref{energ})
occurs using (\ref{G-W}), and yields
\begin{equation}
W_{c}^{n} \simeq \left( \frac{\hbar }{2gV}\right) ^{1/3}\frac{1}{\left(
\sum_{k}1\right) ^{2/3}}.  \label{pot}
\end{equation}%
Note that in the sum over $k^{\prime }$ in (\ref{G-W}) we have approximated the potentials at zero wave vectors. Inserting (\ref{pot}) into (\ref{energ}) we arrive finally for the energy $\mathcal{E}_{0,k}$ of the $k$-mode at
\begin{equation}\label{energy0}
    \mathcal{E}_{0,k} \simeq \frac{9}{2}\frac{\hbar g}{\left(\sum_k1\right)^{2/3}}
    \left(\frac{\hbar}{2gV}\right)^{1/3} .
\end{equation}
Further we replace the sum over $k$ by the integral according to $%
\sum_{k}1\rightarrow V\int d^{3}k/(2\pi )^{3}=V\Lambda^3/6\pi^2$, where the microscopic cutoff $\Lambda $ in the integral over $k$ has been introduced, and obtains in conventional units
\begin{equation}\label{en-0k}
   \mathcal{E}_{0,k} \simeq \frac{9}{4}\left(3\pi^2\right)^{2/3}\frac{ c\hbar}{\Lambda^{2} V}
   \left(g^2/c\hbar\right)^{1/3} .
\end{equation}
The cutoff $\Lambda$ separates the small $k$ from the large ones, where the nonlinearities are expected to be less important.
The multiplication of (\ref{en-0k}) with the factor $V\Lambda^3/6\pi^2$ yields the total vacuum energy for small $k$ as
\begin{equation}
\mathcal{E}_0 \simeq \frac{9}{4\pi 2^{1/3}}\Lambda c\hbar \left( \frac{g^{2}}{c\hbar }%
\right) ^{1/3},  \label{energ2}
\end{equation}
which is independent of the volume. The numerical prefactor for $SU(3)$ is equal to $3\times 9$. Note
that the energy depends non analytically on the fine structure constant.

To get a numerical value of the gap energy for $SU(2)$ we use $\alpha _{w}=g^{2}/c\hbar\simeq 0.004$, set $\Lambda $ equal to the inverse of the interaction radius of weak interactions $\Lambda =m_{W}c^{2}/\hbar c$ with $m_{W}\simeq 80\
GeV/c^{2}$ being the mass of the $W$-boson, and obtain $\mathcal{E}_{SU(2)}\simeq 7.2$ GeV. The estimate for $SU(3)$ with $\alpha _{s}\simeq 1$ and setting $1/\Lambda $ equal to the radius of the proton $\simeq 0.84\times 10^{-15}$ m yields $\mathcal{E}_{SU(3)}\simeq 0.95$ GeV, which is of order of magnitude of the nucleon mass.

To conclude, we have derived using the quantization of Yang-Mills fields via the quantum mechanical equations of motion the non linear commutation relations for the field strengthes of non-Abelian gauge fields at the same time. The equations of motion with respect to the space coordinates (\ref{eq-Espace1},\ref{eq-Bspace1}) coincide with the corresponding classical field equations only by imposing the constraints (\ref{constr_E},\ref{constr_B}). From the commutation relations between the magnetic and electric fields we derive the uncertainty relations for expectation values of vacuum fluctuations of electric and magnetic field strengthes, and use the latter to estimate the energy of vacuum fluctuations of the Non-Abelian gauge fields in the long wave limit. The vacuum energy does not vanish for $k\rightarrow 0$ i.e. it possesses a gap, which is expected to represent a lower bound of the vacuum energy of the Yang-Mills fields.


\begin{thebibliography}{99}
\bibitem{yang-mills54} C. N. Yang and R. Mills, Phys. Rev. \textbf{96}, 191
(1954).

\bibitem{feynman} R.P. Feynman, Acta Physica Polonica, \textbf{24}, 697
(1963).

\bibitem{faddeev-popov67} L.D. Faddeev and V.N. Popov, Phys. Lett. \textbf{B
25}, 29 (1967).

\bibitem{dewitt} B. De Witt, Phys. Rev. \textbf{162}, 1195, 1239 (1967).

\bibitem{dirac-lectures} P. A. M. Dirac, Directions in Physics, John Wiley
\& Sons, New York, 1978.

\bibitem{ytzykson-zuber} C. Ytzykson and J.B. Zuber, Quantum Field Theory,
Vol. 2, MacGraw Hill, New York, 1980.

\bibitem{dirac-qm} P. A. M. Dirac, The Principles of Quantum Mechanics,
Clarendon Press, Oxford, 1958.

\bibitem{landau-lifschitz-bd4} V.B. Berestetskii, E.M. Lifshitz, L.P.
Pitaevskii, Quantum Electrodynamics. Vol. 4 (2nd ed.), Butterworth-Heinemann, 1982.

\end{thebibliography}
\end{document}